# Emotion Recognition In Persian Speech Using Deep Neural Networks

Ali Yazdani, Hossein Simchi, Yasser Shekofteh
Faculty of computer science and engineering (CSE)
Shahid Beheshti University (SBU)
Tehran, Iran
ali.yazdani@mail.sbu.ac.ir, hsimchi74@gmail.com, y_shekofteh@sbu.ac.ir

*Abstract*—Speech Emotion Recognition (SER) is of great importance in Human-Computer Interaction (HCI), as it provides a deeper understanding of the situation and results in better interaction. In recent years, various machine learning and Deep Learning (DL) algorithms have been developed to improve SER techniques. Recognition of the spoken emotions depends on the type of expression that varies between different languages. In this paper, to further study important factors in the Farsi language, we examine various DL techniques on a Farsi/Persian dataset, Sharif Emotional Speech Database (ShEMO), which was released in 2018. Using signal features in low- and high-level descriptions and different deep neural networks and machine learning techniques, Unweighted Accuracy (UA) of 65.20% and Weighted Accuracy (WA) of 78.29% are achieved.

## I. Introduction

Human emotional state is an important factor in their interactions and affects most communication ways such as facial expressions, voice characteristics, and linguistic content of verbal communication [1]–[3]. Speech is one of the main ways to express emotions. To obtain a natural human-computer interaction (HCI), it is highly important to recognize, interpret, and respond to the emotions expressed in the speech [3], [4]. Today, speech emotion recognition (SER) systems have several applications such as human-machine interactions naturally e.g. web videos, computer videos, and training programs, car driver safety, computer games, diagnostic tools to treat the disease, as a tool for an automatic translation system and mobile communications [1]–[3].

Deep learning (DL) has been considered an emerging research field in machine learning (ML) and has received more attention in recent years [5]. DL techniques for SER have several advantages over traditional methods, including their ability to recognize complex structures and features without needing to manually extraction, to extract low-level features from raw data, and the ability to deal with unlabeled data [2], [6]. Our main goal in this paper is to investigate the different speech feature sets and their extraction methods, as well as the impact of using different deep neural network (DNN) architectures to detect the spoken emotions such as anger, surprise, happiness, sadness, and neutral state in the ShEMO dataset, which is a Farsi/Persian spoken speech dataset.

The rest of the structure of this paper is as follows: In section 2, we will review the related works. In section 3, we introduce our contribution which includes reviewing the ShEMO data set and two common methods of extracting features as LLDs (Low-Level Descriptors) and functionals from voice signals, as well as testing different DNN models on these features. In section 4 we will review and discuss the results and finally in section 5, we will have the conclusion.

## II. Related Works

The SER approaches consist of two steps known as feature extraction (FE) and feature classification. At the first step of speech processing, researchers have acquired several characteristics including prosodic and vocal tract features. The second step is about using traditional classifiers such as Support Vector Machine (SVM) or neural networks. Deep neural networks (DNNs) and Convolutional Neural Networks (CNN) provide efficient results for signal processing. On the other hand, recursive networks such as Recursive Neural Networks (RNNs) and Long Short-Term Memory (LSTM) are very effective in SER.

First of all, it should be said that the SER systems can be classified into two main classes, frame-based or utterance-based systems [3], [4], [7], [8]. Input speech signal, as an utterance, is divided into smaller segments, called frames, in many speech processing systems. Frame-based SER systems utilize FE of all of the frames, but utterance-based systems use a fixed size feature vector, known as global or Functional features, after post-processing of the raw features. Using a Bidirectional LSTM (BLSTM) network which effectively preserves the characteristics of temporal dynamics, in [7], on the IEMOCAP database shows that the SER system will perform better than DNN. They obtained low-level acoustic features as well as the global features using statistical functions and then applied them to the output and then the acoustic features are given to the Extreme Learning Machine (ELM) network. Finally, UA was achieved of 63.89% with WA of 62.85%.

In [9], the spectrogram was used directly to train a CNN-LSTM network. At the first step, utterances that are longer than 3 seconds are divided into small equal parts. Also, they calculated spectrogram for each frame by applying Hamming window to each signal with a frame size of 10 milliseconds, a window size of 20 milliseconds, and a Fourier series with a length of 800. In the CNN-LSTM architecture, it is assumed that CNN extracts specific patterns that contain emotional information in the utterance, and LSTM pays attention to the temporal behavior during the utterance. The system without having a separate feature extraction step was achieved WA of 67.3% and UA of 62.0%.

In [8], a BLSTM recursive neural network was used along with the attention mechanism which ignores the silent parts of the utterance as well as the parts that do not have emotional content. Both frame features and temporal aggregation can be learned over longer periods. Therefore, silent frames receive small weights and the rest of the frames receive appropriate weights based on the amount of emotional content. Finally, they achieved WA of 63.5% and UA of 58.8% on the IEMOCAP database. In [4], the FCN (Fully Convolutional Network) is introduced which works with variable length of speech as well as regardless of segmentation. In this paper, a network with convolutional layers with the attention mechanism is used and it works without needing speech segmentation. Using the attention mechanism, the weights of silent frames are completely reduced and are ignored to detect emotions. This model is achieved a WA of 70.4% and UA of 63.9%.

In [10], after converting the signal into spectrogram, the data is transmitted directly to the network. A CNN network is used to extract high-level features and then they have used a recursive network. They also used the vocal tract length perturbation (VTLP) data augmentation technique to overcome the problem of insufficient data. Finally, UA of 65.3% and WA of 66.9% were obtained on the IEMOCAP database.

Also, as we mentioned before, choosing the proper features is an important factor in the SER task. In [11], using the openSMILE tool and emobase2010 feature set, they have extracted a 1582-dimensional vector for each utterance. Using GAN (Generative Adversarial Network) architecture with mixup data augmentation technique, better results in a cross-corpus task have been obtained than previous studies. Also, [12] used the IS09[1] feature set which includes 384 features per utterance, along with SVM and logistic regression. MFCC features were also extracted from a maximum of 120 frames of utterances and a matrix (120, 13) was obtained for each utterance which was used in a stacked LSTM architecture.

III. EXPERIMENTAL SETUP

A. Dataset

The ShEMO dataset in [13], contains 3,000 semi-natural speech files, equivalent to 3 hours and 25 minutes of speech samples collected from online radio broadcasts. These files are in .wav, 16-bit, 44.1 kHz, and single-channel formats. In this dataset, 87 people (including 31 women and 56 men), whose mother tongue is Persian used to examine the 5 main feelings of anger, fear, happiness, sadness, surprise, and neutral state without feeling. 12 persons, including 6 men and 6 women, tag these speech files as emotion labels and are used by voting to determine the final label.

The average duration of utterances is 4.11 seconds with a standard deviation of 3.41. It should be noted that due to the small number of files labeled with fear (a total of 38 files), these files have been removed from the experiments.

B. Features

In general, there are two steps to extract features from audio files. Initially, the audio file is split into smaller speech segments, the frames, and the low-level descriptions (LLDs) are extracted from each frame. Then different statistical functions (such as mean, max, variance, linear regression coefficients, etc.) are applied to the LLDs to obtain a high-level or Functional feature vector for each utterance. Different machine learning models such as SVM or decision tree can be used on Functional feature vectors or neural network models such as recursive or convolutional networks directly using LLDs. Fig. 1, shows the proposed system in this paper, along with the various models we used.

We extracted various LLDs from speech files. For this purpose, we sampled each speech file at a sampling rate of 16KHz and we used it for 7.52 seconds (average total file duration plus standard deviation inspired). Also, we cropped longer files and padded shorter files with zero padding in 7.52 seconds. Then we divided the obtained time series into 32ms as well as 100ms frames, with 50% overlap. Finally, 52 features including various spectral features such as centroid, contrast, roll-off, and bandwidth along with features such as zero-crossing rate and MFFC features are extracted from each frame as Hand Crafted LLDs. In addition, we tested two different feature sets for extracting LLDs using the openSMILE toolkit [14]. It should be noted that these features are extracted from 20ms long frames. The first feature set is called the extended Geneva Minimalistic Acoustic Parameter Set (eGeMAPS) [15], which extracts 25 LLD features (such as MFCC features, frequency-related features, bandwidth, etc.) from each frame. The second feature set is the Interspeech 2016 Computational Paralinguistics Challenge (ComParE_2016) [16], which extracts 65 LLDs per frame.

The openSMILE tool provides various features for extracting from speech files. To extract these features, it splits the speech files into 20ms frames with 10ms overlap, extracts the various LLDs from the frames, and then using several statistical functions on them to achieve a d-dimensional feature vector as Functional features for each utterance. We used 3 different feature sets for our experiments:

- The extended Geneva Minimalistic Acoustic Parameter Set (eGeMAPS): In addition to the LLD features described in the previous section, this feature set includes 88 Functional features such as spectral, frequency, amplitude, and energy.

---

[1] The INTERSPEECH 2009 Emotion Challenge feature set.

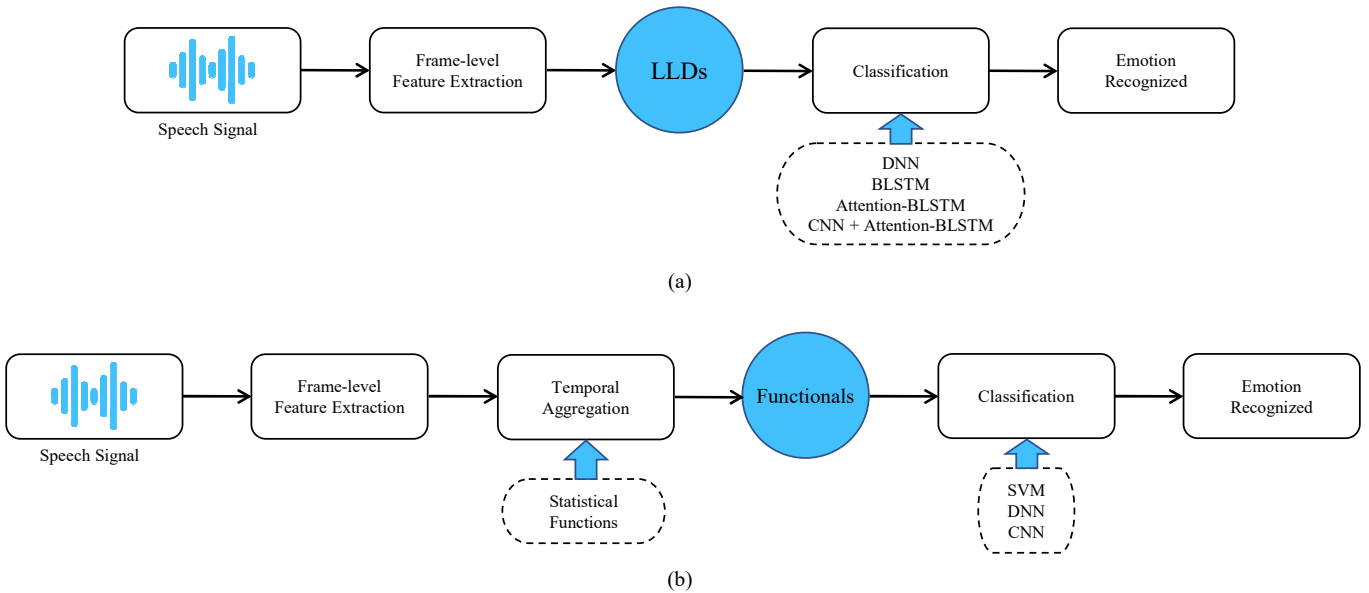

Fig. 1. Two general approaches for developing an SER system: (a) An SER system that uses Low-Level Descriptors extracted from each frame which are frame-level representations. (b) An SER system that uses Functional features obtained by applying various statistical functions to LLDs which results in utterance-level representation.

- The INTERSPEECH 2010 Paralinguistic Challenge feature set (IS10_Paraling) [17]: It uses 21 different statistical functions (such as standard deviation, arithmetic mean, skewness, kurtosis, etc.) to 34 LLD features (including MFCC features, logarithmic power of Mel-frequency bands, and the loudness as the normalized intensity, etc.) is obtained along with their delta. In addition, there are several pitch-related features that ultimately result in a 1582-dimensional feature vector for each utterance.
- The large openSMILE emotion feature set (emo_large): It extracts the largest number of Functional features from each audio file, contains 6552 features obtained by applying more functions to more LLDs.

## IV. DEEP LEARNING MODELS

In recent years, deep learning has been used effectively by researchers due to its multi-layer structure and the presentation of efficient results in a variety of fields, including emotion recognition in speech [1]–[3], [6].

DNNs are based on feed-forward structures, consisting of one or more hidden layers between input and output. CNNs are another type of deep learning technique that is used exclusively with forward-looking architecture for classification. CNNs are commonly used to identify patterns and provide better classification. RNNs are a branch of sequential information neural networks, in which outputs and inputs are interdependent, and their dependence is usually useful in predicting the next state of the input. RNNs, like CNNs, require memory to store general information obtained in a sequential deep learning modeling process, and usually only work efficiently for a few generations. The main problem that affects the overall performance of RNN is its sensitivity to the disappearance of gradients, which leads to forgetting the initial input. To prevent this, LSTM is used to create a block between frequent connections. These networks can also be used Bidirectional-LSTM [2], [5].

The combination of CNN and LSTM networks has received much attention in recent years. In the SER task, it is assumed that CNN extracts specific patterns that contain emotional information in the utterance and LSTM pays attention to the temporal behavior during the utterance [6], [10]. Therefore, using the CNN-LSTM architecture can be effective in categorizing LLD features.

The use of attention mechanisms in neural networks has shown widespread success in a wide range of tasks, such as question answering, machine translation, natural language understanding, and speech recognition. The main idea of the attention mechanism is to focus on a few related parts while ignoring the rest. There are many variations on this mechanism (global vs. local, soft vs. hard) but its main use is to reinforce different LSTM models such as encoder-decoder architecture (for example in machine translation), avoiding the use of a fixed context vector as the only output of the decoder. Specifically, this is the last hidden LSTM layer that carries all the information extracted by the LSTM encoder. Thus, in the classical structure, all information is compressed into a context vector, which can act as a bottleneck, while all hidden middle layers of the encoder are ignored. This vector is then passed to subsequent layers such as the LSTM or dense decoder. In later steps, we rely only on this type of summary given by the encoder, and by increasing the length of the time sequence analyzed, the performance of the model can be reduced. To deal with this problem, the attention mechanism is very effective [2]–[4], [8].

## V. EXPERIMENTS

We tested and evaluated various deep learning models on the ShEMO dataset. Also, we used various LLD as well as Functional acoustic features. The ShEMO dataset is an imbalanced dataset according to the number of files in each class. So, we used both the Unweighted Accuracy (UA) and the Weighted Accuracy (WA) as evaluation metrics. We implemented a 5-fold cross-validation for our experiments. In each fold, four sessions of the data were used to train the model, and the remaining session was used as a test set. The results of our experiments are divided into 3 general sections:

*1) Test Different Models On LLDs:* We first compare the results of applying different deep learning models to the features of Hand Crafted LLDs. In this section, we extract Hand Crafted LLDs in high resolution (32ms frames) and low resolution (100ms frames). All models of deep learning in this section receive inputs (batch size, number of frames, number of features per frame) and after passing different layers, in the end, by the softmax function, the class that is most likely will be determined. The DNN used in the experiment is a Fully Connected Network consisting of dense layers with batch normalization and drop out to prevent overfitting. The use of BLSTM, and BLSTM networks along with the attention mechanism, are two other models that were tested, and finally, a CNN-BLSTM network that uses the attention mechanism was also tested. The architecture of this network is such that it consists of 1-dimensional convolution layers with 1-dimensional max pooling, followed by a Bidirectional LSTM layer that uses the attention mechanism. The results show that the use of the LSTM recursive network on sequential LLD features has improved the model performance over DNN. It is also much more effective than LSTM alone in recognizing emotion in spoken utterances using the attention mechanism associated with BLSTM. Finally, using convolution layers before BLSTM attention, more prominent components in LLDs were extracted first, which improved the performance of the model. It should be noted that this network in low resolution (100ms frames) can not have the same performance as high resolution because, at low resolutions, convolutional layers cannot create effective representations of LLD properties. The results of this section as are given in Table 1, show that using a CNN+Attetnion-BLSTM model can perform well with Hand Crafted LLDs obtained from 32ms frames. This model achieved a UA of 63.52% and WA of 75.32%.

*2) Test CNN+Attention-BLSTM Model On Different LLD Feature Sets:* In previous experiments, we showed that a neural network with a CNN-BLSTM architecture with an attention mechanism can perform well on Hand Crafted LLDs (52 LLDs). In this section, we test this model on the eGeMAPS (25 LLDs) and ComParE_2016 (65 LLDs) feature lists extracted by openSMILE tool. As shown in Table 2, this CNN+Attetnion-BLSTM model with a UA of 63.52% and WA of 75.32%, still performs better on Hand Crafted LLDs than other feature sets.

*3) Test Different Models On Different Functional Feature Sets:* In this section, we test different models of deep learning on Functional features. The neural networks used in this section receive the inputs (batch size, number of features per utterance) and after passing through different layers, like the previous models, the softmax function of the class with the most Specifies the probability. A DNN network consisting of fully connected layers is a CNN network with 1-dimensional convolutional layers. In addition, an SVM model with Radial Basis Function (RBF) kernel was tested for comparison according to the settings used in the ShEMO paper. The results show that using deep neural as well as larger feature sets can have better performance than traditional machine learning models. Compare to the traditional SVM model used in ShEMO paper, +7.0% improvement of UA (from 58.20% to 65.20%) was achieved by using a simple CNN along with emo_large feature set. The feature sets used, the number of their Functional features, models, and the results are listed in Table 3.

TABLE I. COMPARISON OF DIFFERENT MODELS ON LLDS

| Frame Length | Hand Crafted LLDs | | |
|---|---|---|---|
| | *Method* | *UA* | *WA* |
| 100ms | BLSTM | 54.13 | 69.75 |
| | Attention-BLSTM | 60.94 | 74.61 |
| | CNN+Attention-BLSTM | 59.75 | 73.59 |
| | DNN | 51.78 | 64.38 |
| 32ms | BLSTM | 51.04 | 66.91 |
| | Attention-BLSTM | 58.06 | 72.21 |
| | CNN+Attention-BLSTM | **63.52** | **75.32** |
| | DNN | 49.77 | 64.18 |

TABLE II. COMPARISON OF CNN+ATTENTION-BLSTM MODEL ON DIFFERENT LLD FEATURE SETS

| Feature Set | CNN + Attention-BLSTM | |
|---|---|---|
| | *UA* | *WA* |
| ComParE_2016 | 61.00 | 74.40 |
| eGeMAPS | 60.45 | 73.63 |
| Hand Crafted Features | **63.52** | **75.32** |

TABLE III. COMPARISON OF DIFFERENT MODELS ON DIFFERENT FUNCTIONAL FEATURE SETS

| Feature Set | Functionals | | | |
|---|---|---|---|---|
| | *# of features* | *Method* | *UA* | *WA* |
| eGeMAPS | 88 | SVM (baseline) | 58.20 | 73.90 |
| | | DNN | 56.05 | 72.45 |

| Feature Set | Functionals | | | |
| --- | --- | --- | --- | --- |
| | # of features | Method | UA | WA |
| IS10_Paraling | 1582 | CNN | 57.62 | 73.63 |
| | | SVM | 65.02 | **78.29** |
| | | DNN | 64.78 | 76.63 |
| | | CNN | 63.77 | 76.70 |
| emo_large | 6553 | DNN | 64.33 | 76.80 |
| | | CNN | **65.20** | 77.61 |

## VI. CONCLUSION

In this study, we investigated different deep and machine learning techniques, as well as speech features to recognize emotions. Also, due to the novelty of using different deep learning techniques in Farsi, we introduced the best model for recognizing emotions in Farsi, and finally, we were able to obtain higher accuracy than traditional methods. We improved the UA reported in the ShEMO paper by 7%. In addition, we achieved a WA of 78.29%.

The results showed that the use of a Convolutional Neural Network followed by a BLSTM, performs better than a DNN or CNN-only network when using LLDs. Also, the use of the attention mechanism significantly improves the performance of the model in this case. In contrast, when using Functional features, using a machine learning model such as SVM can be a good choice, although when the number of features increases, a CNN network can perform better in terms of computational cost and accuracy.